\documentclass[epj,nopacs]{svjour}
\usepackage{graphics}
\usepackage{graphicx}
\usepackage{feynmf}
\usepackage{sparticles}
\begin{document}
\begin{fmffile}{main}
\onecolumn
\newcommand{\br}{\begin{eqnarray}}
\newcommand{\er}{\end{eqnarray}}
\newcommand{\bea}{\begin{eqnarray}}
\newcommand{\eea}{\end{eqnarray}}
\newcommand{\bi}{\begin{itemize}}
\newcommand{\ei}{\end{itemize}}
\newcommand{\bn}{\begin{enumerate}}
\newcommand{\en}{\end{enumerate}}
\newcommand{\bc}{\begin{center}}
\newcommand{\ec}{\end{center}}
\newcommand{\ul}{\underline}
\newcommand{\ol}{\overline}
\def\epem{\ifmmode{e^+ e^-} \else{$e^+ e^-$} \fi}
\newcommand{\eeww}{$e^+e^-\rightarrow W^+ W^-$}
\newcommand{\qqQQ}{$q_1\bar q_2 Q_3\bar Q_4$}
\newcommand{\eeqqQQ}{$e^+e^-\rightarrow q_1\bar q_2 Q_3\bar Q_4$}
\newcommand{\eewwqqqq}{$e^+e^-\rightarrow W^+ W^-\ar q\bar q Q\bar Q$}
\newcommand{\eeqqgg}{$e^+e^-\rightarrow q\bar q gg$}
\newcommand{\eeqloop}{$e^+e^-\rightarrow q\bar q gg$ via quark loops}
\newcommand{\eeqqqq}{$e^+e^-\rightarrow q\bar q Q\bar Q$}
\newcommand{\eewwjjjj}{$e^+e^-\rightarrow W^+ W^-\rightarrow 4~{\rm{jet}}$}
\newcommand{\eeqqggjjjj}{$e^+e^-\rightarrow q\bar
q gg\rightarrow 4~{\rm{jet}}$}
\newcommand{\eeqloopjjjj}{$e^+e^-\rightarrow q\bar
q gg\rightarrow 4~{\rm{jet}}$ via quark loops}
\newcommand{\eeqqqqjjjj}{$e^+e^-\rightarrow q\bar q Q\bar Q\rightarrow
4~{\rm{jet}}$}
\newcommand{\eejjjj}{$e^+e^-\rightarrow 4~{\rm{jet}}$}
\newcommand{\jjjj}{$4~{\rm{jet}}$}
\newcommand{\qqbar}{$q\bar q$}
\newcommand{\ww}{$W^+W^-$}
\newcommand{\ar}{\rightarrow}
\newcommand{\sm}{${\cal {SM}}$}
\newcommand{\Dir}{\kern -6.4pt\Big{/}}
\newcommand{\Dirin}{\kern -10.4pt\Big{/}\kern 4.4pt}
\newcommand{\DDir}{\kern -7.6pt\Big{/}}
\newcommand{\DGir}{\kern -6.0pt\Big{/}}
\newcommand{\wwqqqq}{$W^+ W^-\ar q\bar q Q\bar Q$}
\newcommand{\qqgg}{$q\bar q gg$}
\newcommand{\qloop}{$q\bar q gg$ via quark loops}
\newcommand{\qqqq}{$q\bar q Q\bar Q$}
\newcommand{\ord}{{\cal O}}
\newcommand{\Ecm}{E_{\mathrm{cm}}}
\setlength{\unitlength}{1mm}
\def \gsim
{\raisebox{-3pt}{$\>\stackrel{>}{\scriptstyle\sim}\>$}}
\title{Higgs and neutrino sector, EDM and $\epsilon_K$ in a
spontaneously CP and R-parity breaking supersymmetric model}
\author{M. Frank\inst{1},  K. Huitu\inst{2} \and T. R\"uppell\inst{2}}
\institute{ Department of Physics, Concordia University,\\
7141 Sherbrooke St. West, Montreal, Quebec, Canada H4B 1R6\\  \and 
High Energy Physics Division, Department of Physical
Sciences, University of Helsinki,\\ and  Helsinki Institute of Physics,
P.O. Box 64, FIN-00014 University of Helsinki, Finland\\ }
\date{Received: date / Revised version: date}
\abstract{We construct  an extension of the supersymmetric standard model
where both CP symmetry and R-parity are spontaneously
broken.  We study the electroweak symmetry breaking sector of the model
and find minima consistent with the experimental bounds on Higgs boson
masses.
Neutrino masses and mixing
angles are generated through both seesaw and bilinear R-parity violation.
We show that the hierarchical mass pattern is obtained,  and mixings are
consistent with measured values.
Due to the spontaneous CP and R-parity violation, the neutrino sector
is CP violating, and we calculate the corresponding phase.
We further restrict the parameter space to agree with the limits on the
electric dipole moment of the neutron.
Finally, we study the CP violation parameter $\epsilon_K$ in the kaon system
and show that we obtain results consistent  with the experimental value.}
\authorrunning{M.~Frank, K.~Huitu, T.~R\"uppell}
\titlerunning{Higgs and neutrino sector in a spontaneous CP and R breaking model}
\maketitle
\section{Introduction}
\label{sec:intro}

While the Standard Model of electroweak interactions (SM) has achieved a
great deal of success, there  still exist fundamental questions for which
it does not provide answers.

One of these challenging questions is charge-parity (CP) violation.
Though at present the Cabibbo-Kobayashi-Maskawa (CKM) mechanism is
in agreement with existing experimental data, it cannot accommodate
the baryon asymmetry of the universe needed in big bang cosmology
\cite{Gavela:1994dt}.
There is thus need for physics coming from beyond
the SM scenarios.
There are two basic possibilities to break CP: explicit at the Lagrangian
level, or spontaneous in the vacuum \cite{Lee:1973iz}.
The SM represents a case in which CP is broken explicitly 
\cite{Weinberg:1976hu},
through the introduction in the Lagrangian of complex Yukawa couplings which
lead to CP violation in the charged-current weak interactions.
Extensions of the SM which introduce new CP violating phases often
lead to phenomenological difficulties.
For instance, a general two Higgs doublet model with acceptable
flavor changing interactions predicts a too large value for
$\epsilon_K$ \cite{Hall:1993ca}.
In supersymmetric (SUSY) extensions of the SM, one has additional sources of
explicit CP violation arising from soft supersymmetry  breaking terms in
the Lagrangian.
While in supersymmetric models a large number of new phases
emerge \cite{Khalil:1999ym}, in a general minimal supersymmetric standard
model (MSSM) these phases are strongly constrained by electric dipole 
moments (EDMs)
\cite{Buchmuller:1982ye}.
Thus, in general, supersymmetric models share the problem of the origin of CP
violation with the Standard Model and generate too large supersymmetric
contributions to the dipole moments (known as the SUSY CP problem) 
\cite{Ellis:1982tk}.

An  alternative scheme which could explain the source of CP violation is
achieved through spontaneous symmetry breaking (SCPV) \cite{Lee:1973iz}.
In this scenario the Lagrangian is invariant under the CP symmetry, but
the ground state is asymmetric and the only sources of CP violation 
are the vacuum phases. Another  motivation for SCPV arises from the 
strong CP problem. In spontaneous CP breaking   ${\bar \Theta}$ 
vanishes at tree level and is calculable at higher orders 
\cite{Branco:1979pv}. Additional justification for studying SCPV 
comes from string theories, where CP exists as a good symmetry, but 
could be broken spontaneously \cite{Strominger:1985it}.
Models with spontaneous CP violation require an extension of the
minimal Higgs structure of the SM \cite{Peccei:1988ci}.
In general, more than one neutral Higgs particle will participate in flavor
changing interactions.
However, these interactions are severely restricted by the smallness 
of the $K_L-K_S$
mass difference.
One must have a mechanism to suppress such contributions either by extending
neutral flavor conservation to the Higgs sector (requiring vanishing of flavor
changing couplings) or requiring the neutral Higgs bosons to be heavy 
\cite{Branco:1999fs}.

Another problem of the Standard Model, which persists in the MSSM, is that
the neutrino masses vanish.
Yet the neutrino experiments have provided strong
evidence for small nonvanishing neutrino masses \cite{Eidelman:2004wy}.
Perhaps the most popular mechanism to explain neutrino masses is the 
seesaw mechanism
\cite{Mohapatra:1979ia,Valle}, which  can generate small masses for neutrinos
by allowing Majorana masses through the introduction of heavy right-handed
neutrinos. Another popular way to explain the neutrino masses is 
through a small violation of
R-parity \cite{Hall:1983id}, $R_p=(-1)^{3B-L+2s}$, where $B$=baryon number,
$L$=lepton number, and $s$=spin of the particle \cite{Santamaria:1987uq}.
Extensive phenomenological studies of R-parity violating effects exist;
for a recent review see \cite{Barbier:2004ez}.

While there is no fundamental reason for the existence of R-parity 
in the MSSM,
it is put in by hand in order to protect the proton from decaying.
However models beyond the MSSM exist, which allow for R-parity violation
in the lepton sector
only.  Such is the case if R-parity is broken spontaneously 
\cite{Masiero:1990uj}.
If only lepton number (or baryon number) is violated, the proton does
not decay.
In,  {\it e.g.},  \cite{Frank:2001tr}, phenomenological implications of  a
model with spontaneous R-parity violation in the quark and leptonic
sectors were explored.

Both neutrino masses and CP violation could be explained if CP and R-parity
were
symmetries of the Lagrangian, but spontaneously broken by the vacuum.
In this paper we consider a model where both of these violations are
intertwined and both are spontaneous.
CP is broken spontaneously through complex vacuum expectation
values (VEVs) of the neutral scalar
bosons, while R-parity violation and the seesaw mechanism will be 
allowed to provide the
required masses and mixing parameters for neutrinos.
We show that while it is non-trivial to satisfy conditions for both
symmetries to
break down at the same time, there are regions in the parameter space where
  suitable Higgs masses, as well as measured neutrino mass
differences and mixing angles are obtained.
In such a model, CP is violated in the neutrino sector.

A model which included both neutrino mass generation mechanisms
- namely seesaw and R-parity violation -
with spontaneous R-parity violation was realized in \cite{Kitano:1999qb},
where  the spontaneous $R_p$ violation
was introduced via a term proportional to $ N L H_2$.
This term represents the familiar bilinear $R_p$ violating term
mixing lepton and Higgs  superfields,
$L H_2$, when the right-handed sneutrino field, $\tilde N$,
develops a vacuum expectation value.
This term, however, also  breaks lepton number  spontaneously, and
thus introduces a superfluous massless Goldstone
boson into the scalar spectrum.
The problem can be solved by adding a singlet $S$ to the
theory, through a term $N^2 S$,
which explicitly breaks lepton number (but not R-parity), if
$S$ is not assigned lepton number~$-2$.

Attempts to violate CP spontaneously, by complex VEVs of the
neutral scalars, exist \cite{Lee:1973iz},
but fulfilling the experimental constraints has proven difficult.
More than one Higgs doublet is needed, see, {\it e.g.},
\cite{Branco:1999fs}.
Spontaneous breaking of CP is not possible at tree level in
the MSSM with two Higgs doublets, nor it is allowed
in a model with
four doublets \cite{Masip:1995sm}.
Studies of  minimal CP violations in the MSSM have shown that if no
other symmetries are imposed, at least two extra singlet fields are
required \cite{Masip:1998ui}.
Instead of adding doublets, or two extra singlets, one can study
extended models, like the NMSSM model of \cite{Nilles:1982dy}, where
the so called $\mu$-problem has been avoided by
adding one singlet and requiring $\mathbf Z_3$ symmetry.
At tree-level one cannot get spontaneous CP violation in this model
either, and consequently radiative corrections must be evoked 
\cite{Ham:2003jf}.
In that case a very light Higgs boson emerges \cite{Georgi:1974au}
as it also happens in the MSSM, if spontaneous CP violation is induced
via radiative corrections \cite{Maekawa:1992un}.
The consequences on the Higgs boson mass were also explored in
\cite{Haba:1995aw}.
Another possibility studied is to discard the $\mathbf Z_3$ symmetry
completely.
On one hand, this way one loses the solution to the $\mu$ problem,
but on the other hand, it is possible to achieve SCPV \cite{Branco:2000dq} and
also solve the problem of domain walls, which
are created during the EW phase transition as the $\mathbf{Z}_3$
symmetry is broken spontaneously.

An interesting model for spontaneous CP violation was presented
in \cite{Hugonie:2003yu}, where the  $\mathbf{Z}_3$ symmetry is replaced
by R-symmetries on the whole superpotential, including non-renormalizable
terms
\cite{Panagiotakopoulos:1998yw}.
The method generates a $\mathbf{Z}_3$ breaking tadpole term for the singlet
field $S$ in the soft SUSY breaking part of the Lagrangian. This tadpole
term
allows for spontaneous CP violation to occur at tree-level
\cite{Masip:1998ui,Hugonie:2003yu}.
The tadpole is assumed to originate from non-renormalizable
interactions, which do not spoil quantum stability. We adopt this 
approach here.

Several models in the literature have explored the consequences of
breaking CP spontaneously.
Some have studied the effects on leptonic observables, such as neutrino
masses in the presence of right-handed neutrinos \cite{Doff:2006rt}
and leptogenesis \cite{Chen:2004ww}; as well as the effects on the
electric dipole moments \cite{Babu:1993qm}.
The consequences of having complex phases in the VEVs
of the Higgs bosons have been analyzed  in the kaon system
\cite{Frampton:1998nc} and B-meson system  \cite{Ball:1999yi}.
The consequence of allowing spontaneous, rather than explicit,
CP breaking, is that the CKM matrix obtained is real.
While several models mentioned above can generate CP violation in the
kaon system that is consistent with the experimental data, a recent
study argues that the CKM matrix is likely complex
\cite{Botella:2005fc};
consequently a ``hybrid" model was constructed in which more than one
source
of CP violation is present,  allowing both a complex CKM matrix and
non-trivial CP phases in the Higgs potential \cite{Branco:2006pj}.
Here we do not study the B-sector in detail and we expect that
modifications in quark sector are needed, \textit{e.g.}, along the lines
discussed above, to fulfill all experimental results.
However, these changes do not qualitatively change the results
obtained in this work, as we discuss later.

Our paper is organized as follows.
We give the Lagrangian  and describe the model we used in Section \ref{model}.
We explore Higgs boson masses and impose the condition that the
masses satisfy experimental
bounds in Section \ref{higgs}.
We show that using both seesaw and R-parity violation, correct neutrino
masses and mixings are obtained, in Section \ref{neutrinomass}.
We also calculate the Jarlskog invariant of the neutrino sector.
We explore the consequences of CP violation in the model and show that we
can obtain a region of the parameter space compatible with the bounds on
the electric dipole moments and obtain the observed
$\epsilon_K$ in the kaon system in Section \ref{edms}.
We conclude in Section \ref{conclusion}.

\section{The Model}
\label{model}

Our model is based on the superpotential
\begin{eqnarray}
W &=& \varepsilon_{\alpha\beta}\left(h_U^{ij}Q_i^\alpha H_2^\beta U_j
    +h_D^{ij}H_1^\alpha Q_i^\beta D_j
+h_E^{ij}H_1^\alpha L_i^\beta E_j+h_N^{ij}L_i^\alpha H_2^\beta N_j
    +\lambda_H H_1^\alpha  H_2^\beta S\right)
   +{\lambda_S \over 3!} S^3
    +{\lambda_{N_i} \over 2} N_i^2 S,
\end{eqnarray}
where $H_1$ and $H_2$ denote the Higgs doublet superfields, $L_i$ and $Q_i$
the left-handed lepton and quark doublet superfields,
respectively, and $E_i$ and $U_i$, $D_i$ the lepton and quark singlet
superfields.
Right-handed neutrino superfields are denoted by $N_i$, $i=1,2,3$,
and $S$ is the
gauge singlet superfield. The $SU(2)$ contraction is defined as
$\varepsilon_{12}=-\varepsilon_{21}=1$.
Assuming baryon number conservation and $\mathbf Z_3$ symmetry,
the terms in the superpotential are the only renormalizable ones
that respect CP and R-parity, in addition to the gauge
symmetry. All the parameters in the Lagrangian are real.

The soft SUSY breaking terms in this model are the mass terms for
scalars and gauginos, trilinear $A$-terms, and the additional $S$-tadpole,
\begin{eqnarray}
-V_{soft}&=&M_Q^{ij~2}{\tilde Q}_i^{\ast}{\tilde Q}_j
+M_U^{ij~2}{\tilde U}_i^{\ast}{\tilde U}_j + M_D^{ij~2}{\tilde 
D}_i^{\ast}{\tilde D}_j
  +M_L^{ij~2}{\tilde L}_i^{\ast}{\tilde L}_j +
M_E^{ij~2}{\tilde E}_i^{\ast}{\tilde E}_j
  \nonumber \\
&+&M_N^{ij~2}{\tilde N}_i^{\ast}{\tilde N}_j
+ M_S^2S^{\ast} S
+m_{H_1}^2H_1^{\ast}H_1 + m_{H_2}^2H_2^{\ast}H_2 \nonumber \\
&-& \frac12 (M_3\tilde g \tilde g +M_2\tilde W \tilde W +M_1\tilde B
\tilde B  + \mathrm{h.c.}) \nonumber \\
&+& \bigg[\varepsilon_{ab}(A_U^{ij}\tilde{Q}_i^a\tilde{U}_jH_2^b
+A_D^{ij}\tilde{Q}_i^a\tilde{D}_jH_1^b+
A_E^{ij}\tilde{L}_i^a\tilde{E}_jH_1^b
+A_N^{ij}\tilde{L}_i^a\tilde{N}_jH_2^b
+A_H H_1^a H_2^bS ) \nonumber \\
&+& \sum_i {A_{N_i} \over 2} S\tilde{N}_i^2
+ {A_{S}\over 3!} S^3
+\xi^3 S +\mathrm{h.c.}\bigg].
\end{eqnarray}
Here $i,j$ run over the family indices.
In the tadpole term, $\xi^3 S$, the parameter $\xi$, which originates
from nonrenormalizable terms \cite{Hugonie:2003yu}, has been taken to
be a free parameter, of the order of the soft supersymmetry breaking
terms.
We impose a flavor diagonal texture on $h^{ij}_{U,D,E}$ and the
corresponding $A$-terms.
The full tree-level scalar potential is $V_s=V_{soft}+V_F+V_D$,
where $V_F$ and $V_D$ are the usual $F$ and $D$ terms.
All together, the model contains in the superpotential
51 additional parameters compared to the general MSSM.
If we had instead the MSSM with additional explicit R-parity
violating terms, there would be 48 new couplings
\cite{Barbier:2004ez}.
If in addition CP would be broken, a large number of phases would
appear in the soft masses and couplings.
Compared to these numbers of parameters, our model with
spontaneous CP and R-parity violation is economical.

The minimization of the scalar potential with respect to the moduli
of the scalar fields $\phi_i$ and the corresponding phases $\theta_i$
yield constraints later used in finding the scalar mass matrix,
\begin{eqnarray}
\label{eqn:vac}
\left. {\partial V_s \over
\partial\phi_i}\right|_{\phi=\langle\phi\rangle}
  = 0,\;\;
  \left. {\partial V_s \over
\partial\theta_i}\right|_{\phi=\langle\phi\rangle}
= 0.
\end{eqnarray}

Without spontaneous CP violation, the VEVs are real and the
minimization equations with respect to the phases are always satisfied.
The minimization equations for the charged scalars can be trivially
solved by setting all charged scalar VEVs to zero.
As long as the tree-level masses of these fields
remain positive and the corresponding soft $A$-terms remain small
enough, this is also the global minimum of the potential
with respect to these fields \cite{Kitano:1999qb}.
The complex VEVs remain free parameters and we denote
(the phase of $H_1$ can always be rotated away):
\begin{eqnarray}
\nonumber
&&\langle H_1 \rangle = \left(\begin{array}{c}v_1\\0\end{array}\right),\;
\langle H_2 \rangle =  \left(\begin{array}{c}0\\v_2
     e^{i\delta_2}\end{array}\right), \;
\langle S \rangle = \sigma_S e^{i\theta_S},\\
&&\langle \tilde L_i \rangle =  \left(\begin{array}{c}\sigma_{L_i}
     e^{i\theta_{L_i}}\\0\end{array}\right),\;
\langle \tilde N_i \rangle = \sigma_{R_i} e^{i\theta_{R_i}}.
\end{eqnarray}
Note that since $R_p$ is violated, the $W$ mass is $m_W^2= {1\over 2}
g_2^2 v^2$
where $ v^2 \equiv v_1^2
+ v_2^2 + \sigma_{L_i}^2 \approx (174$ GeV$)^2$.

Minimizing with respect to the neutral scalars, we get at the tree-level:
\begin{eqnarray}
\nonumber
{\partial V\over \partial v_1} &=&
2 v_1 (m_{H_1}^2
        + \lambda_H^2 (v_2^2 + \sigma_S^2)
        + {g_1^2 + g_2^2 \over 4}(v_1^2-v_2^2+\sigma_{L_i}^2) \\
&+& \tan\beta ( {1\over 2}\lambda_H \lambda_S \sigma_S^2
\cos(\delta_2-2\theta_S)
+ {1\over 2}\lambda_H \lambda_{N_i} \sigma_{R_i}^2 \cos(\delta_2+
2\theta_{R_i})
+ A_H \sigma_S \cos(\delta_2+\theta_S) ) ) \nonumber\\
&+& 2 \lambda_H h_N^{ij}\sigma_S\sigma_{L_i}\sigma_{R_j}\cos(\theta_S
-\theta_{L_i}+\theta_{R_j})\nonumber ,\\
\nonumber
{\partial V\over \partial v_2} &=&
2 v_2 (m_{H_2}^2
        + \lambda_H^2 (v_1^2 + \sigma_S^2)
        - {g_1^2 + g_2^2 \over 4}(v_1^2-v_2^2+\sigma_{L_i}^2) \\
\nonumber
&+& h_N^{ij}h_N^{ik} (\sigma_{R_j} \sigma_{R_k}
\cos(\theta_{R_j}-\theta_{R_k})
+ \sigma_{L_j} \sigma_{L_k} \cos(\theta_{L_j}-\theta_{L_k})) \nonumber\\
\nonumber
&+& \cot\beta ( {1\over 2}\lambda_H \lambda_S \sigma_S^2 \cos(\delta_2
-2\theta_S)
+ {1\over 2}\lambda_H \lambda_{N_i} \sigma_{R_i}^2
\cos(\delta_2+2\theta_{R_i})
+ A_H \sigma_S \cos(\delta_2+\theta_S) ) ) \\
&+& 2 A_N^{ij} \sigma_{L_i} \sigma_{R_j} \cos(\delta_2+\theta_{L_i}
-\theta_{R_j})
+ 2 h_N^{ij}\lambda_{N_j}\sigma_S\sigma_{L_i}\sigma_{R_j}\cos(\delta_2
-\theta_S+\theta_{L_i}+\theta_{R_j})\nonumber ,\\
\nonumber
{\partial V\over \partial \sigma_S} &=&
2 \sigma_S (m_S^2
            + \lambda_H^2(v_1^2+v_2^2)
            + \lambda_S \lambda_H v_1 v_2 \cos(\delta_2-2\theta_S)
            + \lambda_{N_i}^2 \sigma_{R_i}^2 \\ \nonumber
            &+& {1\over 2}A_S \sigma_S \cos(3\theta_S)
            + {1\over 2}\lambda_S^2 \sigma_S^2
+ {1\over 2}\lambda_S \lambda_{N_i}\sigma_{R_i}^2\cos(2\theta_S
+2\theta_{R_i}))\nonumber\\ \nonumber
&-&2\xi^3 \cos(\theta_S)
+ 2 A_H v_1 v_2 \cos(\delta_2+\theta_S)
+ A_{N_i} \sigma_{R_i} \cos(\theta_S -2\theta_{R_i})\\
&+& 2 h_N^{ij} \sigma_{L_i}\sigma_{R_j} (\lambda_H v_1 \cos(\theta_S
-\theta_{L_i}+\theta_{R_j})
+\lambda_{N_j} v_2 \cos(\delta_2-\theta_S-\theta_{L_i}+\theta_{R_j}))
\nonumber ,\\
\nonumber
{\partial V \over \partial \sigma_{L_i}} &=&
2 \sigma_{L_i} (M_{L_i}^2
+ {g_1^2 + g_2^2 \over 4}(v_1^2-v_2^2+\sigma_{L_j}^2) ) \nonumber
\\ \nonumber
&+& 2 A_N^{ij} v_2 \sigma_{R_j} \cos(\delta_2+\theta_{L_i}-\theta_{R_j})
+ 2 \lambda_H h_N^{ij} v_1 \sigma_S \sigma_{R_j}
\cos(\theta_S-\theta_{L_i}
+\theta_{R_j})\\ \nonumber
&+& 2  h_N^{ik}h_N^{jk} v_2^2 \sigma_{L_j}
\cos(\theta_{L_i}-\theta_{L_j})
+ 2 h_N^{ij} \lambda_{N_j} v_2 \sigma_S \sigma_{R_j}
\cos(\delta_2-\theta_S+\theta_{R_j}
+\theta_{L_i})\\
&+& 2
h_N^{ij}h_N^{kl}\sigma_{R_j}\sigma_{L_k}\sigma_{R_l}\cos(\theta_{L_i}
-\theta_{L_k}+\theta_{R_l}-\theta_{R_j})\nonumber ,\\
{\partial V \over \partial \sigma_{R_i}} &=&
2 \sigma_{R_i} (M_{R_i}^2
+ A_{N_i} \sigma_S \cos(\theta_S-2\theta_{R_i})
+ \lambda_{N_i} \lambda_H v_1 v_2 \cos(\delta_2+2\theta_{R_i})\nonumber\\
\nonumber
&+&\lambda_{N_i} \lambda_S \sigma_S^2 \cos(2\theta_S+2\theta_{R_i})
                + \lambda_{N_i}^2 \sigma_S^2
+ \lambda_{N_i} \lambda_{N_j}\sigma_{R_j}^2 \cos(2\theta_{R_i}
-\theta_{R_j}))\\ \nonumber
&+& 2 h_N^{ji}h_N^{jk} v_2^2 \sigma_{R_k}
\cos(\theta_{R_i}-\theta_{R_k})
+ 2 A_N^{ji} v_2 \sigma_{L_j} \cos(\delta_2
-\theta_{R_i}+\theta_{L_j})\\
\nonumber
&+& 2 \lambda_H h_N^{ji} v_1 \sigma_S \sigma_{L_j}
\cos(\theta_S+\theta_{R_i}
-\theta_{L_j})
+ 2 \lambda_{N_i} h_N^{ji} v_2 \sigma_S \sigma_{L_j}
\cos(\delta_2-\theta_S
+\theta_{R_i}+\theta_{L_j})\\
&+& 2 h_N^{ji}h_N^{lk} \sigma_{L_j} \sigma_{R_k} \sigma_{L_l}
\cos(\theta_{R_i}-\theta_{R_k}+\theta_{L_l}-\theta_{L_j})
\nonumber ,\\
\nonumber
{\partial V \over \partial \delta_2}&=&
\lambda_H v_1 v_2 (-\lambda_S \sigma_S^2 \sin(\delta_2 - 2 \theta_S)
- \lambda_{N_i} \sigma_{R_i}^2 \sin(\delta_2 + 2 \theta_{R_i}))
- A_H v_1 v_2 \sigma_S \sin(\delta_2+ \theta_S) \nonumber \\
&-& 2 v_2 ( \lambda_{N_j} h_N^{ij} \sigma_S \sigma_{R_j} \sigma_{L_i}
\sin(\delta_2 - \theta_S + \theta_{R_j} + \theta_{L_j})
+ A_N^{ij} 
\sigma_{R_j}\sigma_{L_i}\sin(\delta_2-\theta_{R_j}+\theta_{L_i})) 
\nonumber ,\\
{\partial V \over \partial \theta_S}&=&
2  \sigma_S (\lambda_H \lambda_S v_1 v_2 \sigma_S \sin(\delta_2-2\theta_S)
- A_H v_1 v_2 \sin (\delta_S + \theta_S)
+  \xi^3 \sin(\theta_S) \nonumber\\
&-& {1\over 2}A_S \sigma_S^2 \sin(3 \theta_S)
- {1\over 2}A_{N_i} \sigma_{R_i}^2\sin(\theta_S-2\theta_{R_i})
- {1\over 2}\lambda_S \lambda_{N_i} \sigma_S \sigma_{R_i}^2 
\sin(2\theta_S + 2\theta_{R_i})
\nonumber\\
&-& \lambda_H h_N^{ij} v_1 \sigma_{R_j}\sigma_{L_i} \sin(\theta_S + 
\theta_{R_j} -\theta_{L_i})
+\lambda_{N_j} h_N^{ij} v_2 \sigma_{R_j} \sigma_{L_i}
\sin(\delta_2 -\theta_S +\theta_{R_j} +\theta_{L_i})  )\nonumber ,\\
{\partial V \over \partial \theta_{R_i}}&=& 2 \sigma_{R_i} 
(\sigma_{R_i}(A_{N_i} \sigma_S
\sin(\theta_S -2\theta_{R_i})
-\lambda_H v_1 v_2  \sin(\delta_2 + 2 \theta_{R_i})
-{1\over 2}\lambda_S \sigma_S^2
\sin(2\theta_S + 2 \theta_{R_i}))\nonumber\\
&-&\sigma_S \sigma_{L_j} h_N^{ji}(\lambda_H v_1 \sin(\theta_S 
+\theta_{R_i}-\theta{L_j})
+\lambda_{N_i} v_2 \sin(\delta_2 -\theta_S +\theta_{R_i} 
+\theta_{L_j}))\nonumber\\
&+&A_N^{ji} v_2 \sigma_{L_j} \sin(\delta_2 - \theta_{R_i} +\theta_{L_j})
+ h_N^{ji}h_N^{jk} v_2^2 \sigma_{L_k} 
\sin(\theta_{R_k}-\theta_{R_i})\nonumber\\
&+&{1\over 2}\lambda_{N_i}\lambda_{N_j} \sigma_{R_i} \sigma_{R_j}^2
\sin(2\theta_{R_j}-2\theta_{R_i})
+h_N^{ji}h_N^{kl} \sigma_{R_l}\sigma_{L_j}\sigma_{L_k}
\sin(\theta_{R_l}-\theta_{R_i}+\theta_{L_k}-\theta_{L_j}))\nonumber ,\\
{\partial V \over \partial \theta_{L_i}}&=& 2 \sigma_{L_i} (\sigma_S 
\sigma_{R_j} h_N^{ij}
(\lambda_H v_1\sin(\theta_S +\theta_{R_j}-\theta_{L_i}) - \lambda_{N_j} v_2
\sin(\delta_2 -\theta_S+\theta_{R_j}+\theta_{L_i}))\nonumber\\
&-&A_N^{ij} v_2 \sigma_{R_j} \sin(\delta_2 -\theta_{R_j}+\theta_{L_i})
+h_N^{ij}h_N^{kj} v_2^2 \sigma_{L_k} \sin(\theta_{L_k}-\theta_{L_i})
\nonumber\\
&+& h_N^{ij}h_N^{kl} \sigma_{L_k}\sigma_{R_j}\sigma_{R_l}
\sin(\theta_{L_k}-\theta_{L_i}+\theta_{R_l}-\theta_{R_j})) .
\end{eqnarray}
Here $g_1$ and $g_2$ are the U(1) and SU(2) gauge couplings,
respectively, and $\tan\beta =v_2/v_1$.
We use the seventeen minimization equations above
solve for the
soft masses of the neutral scalar fields and a subset of $A$-parameters
($A_H,\,A_S,\,A_{N_i},\,A_N^{i,3}$).

\section{Higgs Masses}
\label{higgs}

Separating the real and imaginary parts ($\phi \equiv \phi_r + i
\phi_i$) of the nine neutral scalar fields
(two Higgs, one singlet and six sneutrinos) we get
an 18$\times$18 dimensional mass matrix for the scalars.
The radiative corrections to the scalar masses are implemented
via the one-loop effective scalar potential \cite{Brignole:1991pq},
\begin{eqnarray}
V_{1-loop}&=&{-3\over 32 \pi^2}\left[
\sum_{\tilde{a}=1}^4 m^4_{\tilde a}\left(\log{m^2_{\tilde a}\over
\Lambda^2}-{3\over 2}\right)
-\sum_{{a}=1}^4m^4_a\left(\log{m^2_a\over\Lambda^2}
-{3\over 2}\right)\right],
\end{eqnarray}
where $m_{\tilde a}^2$ are the field dependent eigenvalues of the
$4\times 4$ $\tilde b$-$\tilde t$ mass matrix and
$m_t^2=y_t^2 v_2^2$, $m_b^2=y_b^2 v_1^2$.
$\Lambda$ is the renormalisation scale.
The loop corrections lead to additional terms in both the
minimisation conditions and the scalar mass matrix.  In numerical
calculations, we omit the D-term contributions and set for simplicity,
$M_{Q_{33}}=M_{U_{33}}=M_{SUSY}$, with $M_{SUSY}\sim \Lambda\sim 1$ TeV.

The following experimental input is used:
\begin{eqnarray} \nonumber
&& v = 174\; \mathrm{ GeV},\;
m_W = 80.42\; \mathrm{ GeV},\;
m_t^{Pole} = 175\; \mathrm{ GeV},\;\nonumber\\
&&\alpha_s = 0.102,\;
m_\tau = 1.777\; \mathrm{ GeV},\; \sin^2\theta_w = 0.23124.
\end{eqnarray}
Here
\begin{equation}
m_t  = {m_t^{Pole}\over 1+ {4\over 3\pi}\alpha_s}.
\end{equation}
The rest of our free parameters are randomly sampled, with
sampling ranges as follows (the couplings $\lambda_i$ are constrained by
perturbativity):
\bea
&&0.1<\lambda_{H,N_i}<0.4,\;\;0.2<\lambda_S<0.7,\;\;
|h_N|<10^{-7},\nonumber\\
&&0.4 \;{\rm TeV}< \xi <1\;{\rm TeV},\;\;
-\pi<\theta_{\phi}<\pi,\;\;
|\langle S\rangle|<1\;{\rm TeV},\;\;\nonumber\\
&&|\langle \tilde\nu_L\rangle|<100\;{\rm keV},\;\;
|\langle \tilde N_i\rangle|<1\;{\rm TeV},\;\;
2<\tan\beta<60,
\eea
and the $A$-parameters not eliminated by Eq. (\ref{eqn:vac}) vary between
$0<A_N^{ij}<(1\;{\rm TeV}) h_N^{ij}$.
It should be noted that the VEVs of the right-handed sneutrinos are not
constrained by any experimental bounds. In principle $\tilde N$
could develop a VEV at a different scale than all the other scalars. However,
if this VEV were to be complex, it would propagate CP
violating phases to other (real valued) parameters of our model via their
RG running equations and thus we restrict $\sigma_R\simeq M_{SUSY}$.

\begin{figure}[tpb]
\begin{center}
\includegraphics[width=8truecm,height=6truecm]{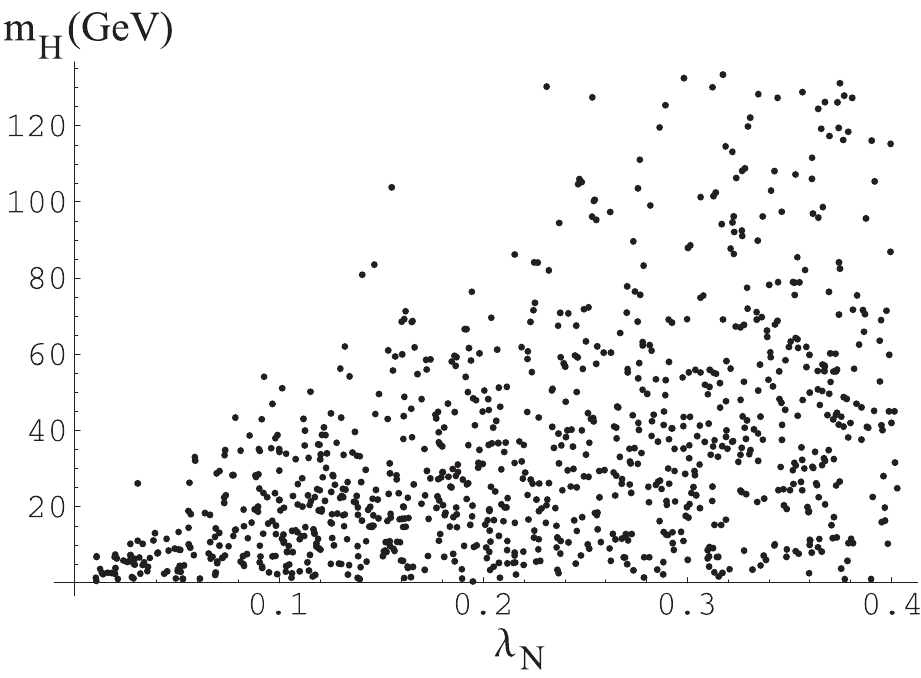}
\end{center}
\caption{
Lightest scalar mass as a function of $\lambda_N$.
}
\label{Goldstone}
\end{figure}

In the limit $\lambda_{N_i} \to 0$ we recover lepton number conservation.
CP and R-parity are still spontaneously violated as is lepton number,
and thus in this limit  the neutral scalar spectrum contains an
additional Goldstone, as illustrated in Fig. \ref{Goldstone}.
Using a model with only one right-handed neutrino (and consequently only
one $\lambda_N$), the
lightest scalar mass clearly tends to zero as the coupling $\lambda_N
\to 0$.

For the full model we choose to set
$\sigma_{R_{1,2}}=\theta_{R_{1,2}}=0$ to further reduce the sampling
space. This choice identically solves
the vacuum conditions $\partial_{\theta_{R_1}}V=0$ and
$\partial_{\theta_{R_2}}V=0$.
\begin{figure}[tpb]
\begin{center}
\includegraphics[width=9truecm, height=7truecm]{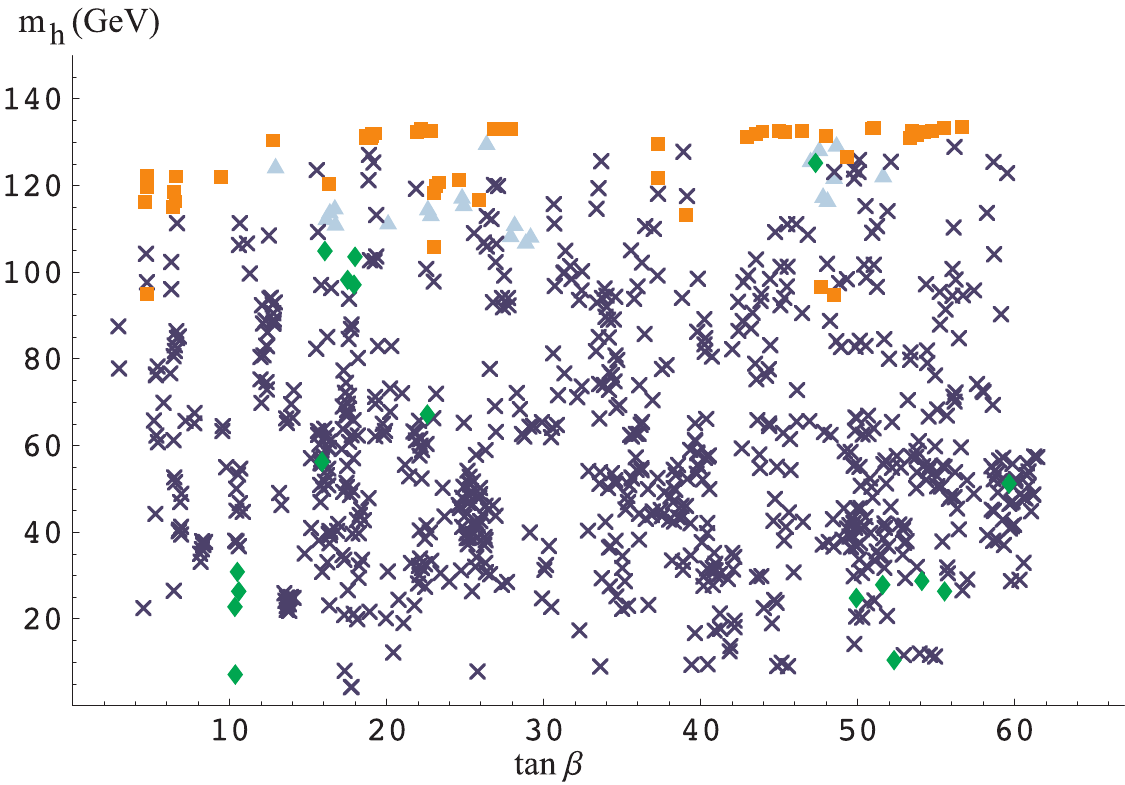}
%
\end{center}
\caption{
The lightest physical scalar mass as a function of $\tan\beta$.
The main component of the lightest scalar is $H_{1,2}^0$ (orange
rectangle), $S$ (green diamond), $\tilde\nu_L$ (light blue triangle),
or $\tilde N$ (dark blue cross).}
\label{Higgs}
\end{figure}
\begin{figure}[tpb]
\begin{center}
\includegraphics[width=10truecm, height=7.7truecm]{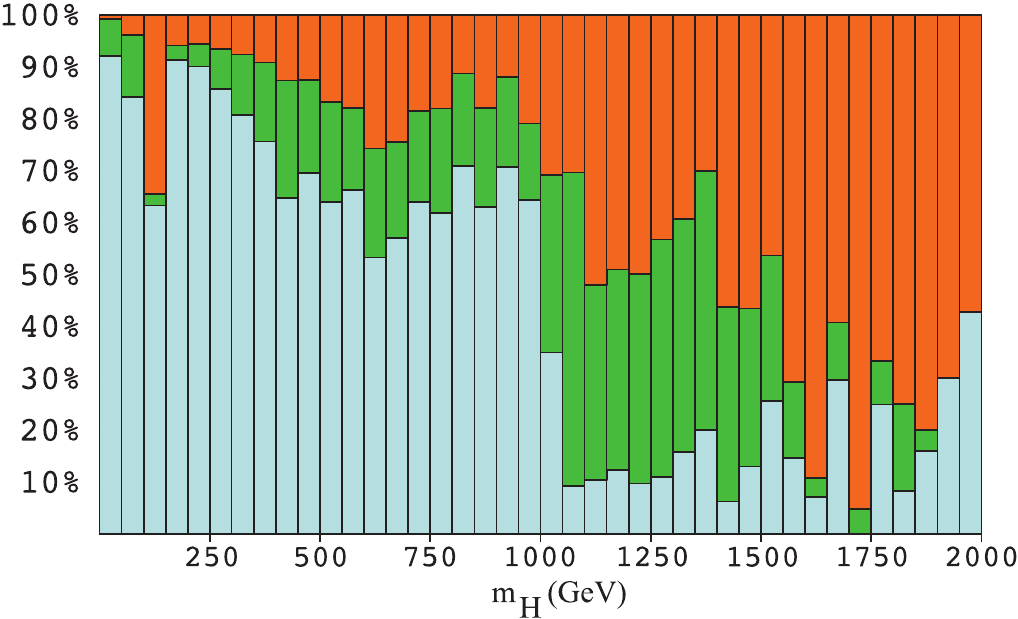}
%
\end{center}
\caption{
Average composition of neutral Higgs particle as a function of
the Higgs mass.
Lightest grey/blue correspond to the sneutrinos, medium grey/green
to singlet, and the darkest grey/orange to doublet Higgs.
}
\label{Higgs_composition}
\end{figure}
\begin{figure}[tpb]
\begin{center}
\includegraphics[width=10truecm ,height=7.7truecm]{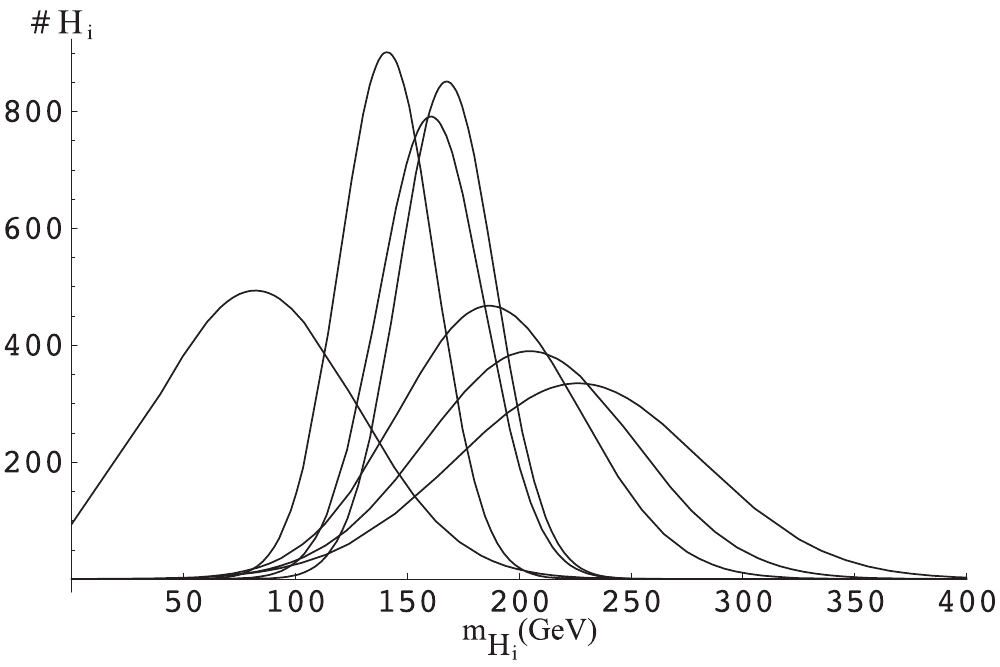}
%
\end{center}
\caption{
Masses of the seven lightest neutral Higgses.  The first peak
corresponds to the
lightest Higgs, second to second lightest etc.}
\label{Higgs_peaks}
\end{figure}
The $ZZh_i$ couplings ($h_i$ denotes any neutral Higgs)
are reduced compared
to the SM, due to other than Higgs doublet components in the
physical Higgs bosons.
Thus the Higgs strahlung production is less frequent than in
the SM and experimental bound on the lightest
Higgs mass is reduced from the SM value of $m_H \gsim 114$ GeV
\cite{lepsearches}, see, \textit{e.g.}, discussion in \cite{Davies:2001uv}.
Similarly the rate for the associated production of two Higgses
through the couplings $Zh_i h_j$ is reduced.
In Fig. \ref{Higgs} we plot the mass of the
lightest Higgs boson as a function of $\tan\beta$.
In the figure we have applied the experimental limits from LEP
\cite{lepsearches2}
on all neutral spin-0 particles to check that the masses are
acceptable and 
we indicate the dominant component of the lightest Higgs.
We have also studied the masses of the charged scalars at
one-loop level
and applied the experimental limits according to the main component
of the charged scalar, {\it i.e.} if the main component is
stau, we have applied the experimental limit of 81.9 GeV, as
appropriate for stau \cite{Yao:2006px}.
Since the charged scalar can be relatively light, it is interesting
to consider the possibility of seeing it at Tevatron in the decay
$t\rightarrow H^+b$.
It appears that the light charged scalars are, however,
mostly sleptons.
Thus the coupling to quarks may be too weak to produce a
significant branching ratio.

In Fig. \ref{Higgs_composition} the composition of all the
neutral Higgses is depicted as a function of their masses for
one thousand parameter points satisfying the constraints
mentioned above.
It is seen that the light experimentally allowed Higgses
tend to be mostly
sneutrinos.
In the region 100-150 GeV,  a significant doublet Higgs component
appears, as can also be seen from Fig. \ref{Higgs}
showing that the lightest neutral Higgs is most often mostly
doublet, if its mass is above 110 GeV.
The heavier Higgses are mostly either singlets or doublets.
In Fig. \ref{Higgs_peaks} a Gaussian is fit to the number
vs mass of the seven lightest neutral Higgses for one thousand
parameter points.
Curves for several of the lightest Higgses are strongly peaked,
showing strong preference for particular mass values.
The curves for heavier Higgses are much broader, showing
much larger variation in their masses.
Interestingly more than half a dozen are in the mass reach of
the LHC.
Unfortunately the doublet component in the light Higgses tends
to be small, as seen in the Fig. (\ref{Higgs_composition}) and
their detection at LHC may be challenging.
Detailed study of the detection is beyond the present work.

\section{Neutrino Masses and Mixing}
\label{neutrinomass}

Since the scalars VEVs appearing in the neutrino mass matrix include
phases, it is expected that the neutrino sector of the model is
CP violating.

In a field basis of $\nu_{L_i},N_i,\tilde S,
\tilde H_1^0, \tilde H_2^0, \tilde B, \tilde W$ the neutral fermions
form the following 11$\times$11 mass matrix:
\begin{equation}
M_{\chi^0} = \left(\begin{array}{ccccccc}
\mathbf{0}_{3\times 3} &
\mathbf{h}_N^{3\times3} \langle H_2^0\rangle &
\mathbf{0}_{3\times 1} &
\mathbf{0}_{3\times 1} &
h_N^{i,j} \langle\tilde N_{j}^\ast\rangle &
- {g_1\over \surd 2} \langle\tilde\nu_{L_i}^\ast\rangle &
{g_2\over \surd 2} \langle\tilde\nu_{L_i}^\ast \rangle \\
\mathbf{h}_N^{3\times 3} \langle H_2^0\rangle &
\mathbf{1}_{3\times 3}\lambda_{N_i} \langle S\rangle &
\lambda_{N_i} \langle\tilde N_{i}^\ast\rangle &
0 &
h_N^{j,i}\langle\tilde\nu_{L_j}\rangle &
0 &
0 \\
\mathbf{0}_{1\times3} &
\lambda_{N_i} \langle\tilde N_{i}^\ast\rangle &
\lambda_S \langle S\rangle &
\lambda_H \langle H_2^0\rangle &
\lambda_H \langle H_1^0\rangle &
0 &
0 \\
\mathbf{0}_{1\times3} &
0 &
\lambda_H  \langle H_2^0\rangle &
0 &
\lambda_H \langle S\rangle &
- {g_1\over \surd 2}  \langle H_1^0\rangle &
{g_2\over \surd 2} \langle H_1^0\rangle \\
h_N^{i,j} \langle\tilde N_{j}^\ast\rangle &
h_N^{j,i}\langle\tilde\nu_{L_j}\rangle  &
\lambda_H \langle H_1^0\rangle &
\lambda_H \langle S\rangle &
0 &
  {g_1\over \surd 2}  \langle H_2^{0\ast}\rangle &
- {g_2\over \surd 2} \langle H_2^{0\ast}\rangle \\
- {g_1\over \surd 2} \langle\tilde\nu_{L_i}^\ast\rangle &
0 &
0 &
- {g_1\over \surd 2} \langle H_1^0\rangle &
{g_1\over \surd 2} \langle H_2^{0\ast}\rangle &
M_1 & 0 \\
{g_2\over \surd 2} \langle\tilde\nu_{L_i}^\ast\rangle &
0 &
0 &
{g_2\over \surd 2} \langle H_1^0\rangle &
- {g_2\over \surd 2} \langle H_2^{0\ast}\rangle &
0 &
M_2
\end{array}\right).
\label{numass}
\end{equation}
The mass matrix Eq. (\ref{numass}) differs from the neutrino mass
matrix in \cite{Kitano:1999qb} by the phases for the VEVs.
It is easy to see the structure of the usual seesaw mechanism, which
produces small neutrino masses $m_\nu$,
\begin{equation}
M_{\chi^0} = \left(\begin{array}{cc}
0   & m_D\\
m_D^T & M_R \end{array}\right), \quad m_\nu=
-m_D M_R^{-1} m_D^T,
\end{equation}
where $ m_D \ll M_R$. Similarly to \cite{Kitano:1999qb},
there are actually several sources for neutrino masses: the usual seesaw and
the mixing of neutrinos with $\tilde{h}_2^0$ and gauginos through 
R-parity breaking.

It can be shown that
the number of independent vectors in $m_D$ is an upper bound of the number of
non-zero neutrino masses, {\it e.g.}, in models with exclusively the 
gaugino seesaw,
  there is at most one nonzero neutrino mass at tree-level because
  there is only one linearly independent vector, $\langle \tilde
  \nu_{L_i}^\ast \rangle$, in $m_D$.
 From (\ref{numass}) it is immediately apparent that we have four 
independent vectors in
$m_D$: three in the Yukawa matrix $\mathbf{h}_N^{3\times3}$, and the
vector of sneutrino vevs $\langle\tilde\nu_{L_i}^\ast\rangle$.
If we were to include only one right-handed
neutrino, there would be two linearly independent vectors,
and as expected, we find that in such models there is one massless neutrino.
In models with no right-handed neutrinos but bilinear R-parity violation,
there are two independent vectors,
$\mu_i$ ($\equiv h_N^{i,j} \langle\tilde N_{j}^\ast\rangle$ in our
model) and $\langle\tilde\nu_{L_i}^\ast\rangle$. The latter, however, can be
rotated away using the accidental $SU(4)$ symmetry of the \{$L_i,H_1$\}
fields, leaving only one independent vector and thus two massless
neutrinos.

Inspecting the requirement that $m_D\ll M_R$ yields some qualitative
features of the model. In particular,
the left-handed sneutrino VEVs must be
  small and $h_N \langle \tilde N^\ast\rangle$ should be of the same
order. Thus, although $\langle \tilde N^\ast\rangle$
is not bound by any other prior consideration, having $h_N\approx
10^{-7}$ results in an upper limit of a few TeV
for the right-handed sneutrino VEVs.

\begin{figure}[tpb]
\begin{center}
\includegraphics[width=10truecm]{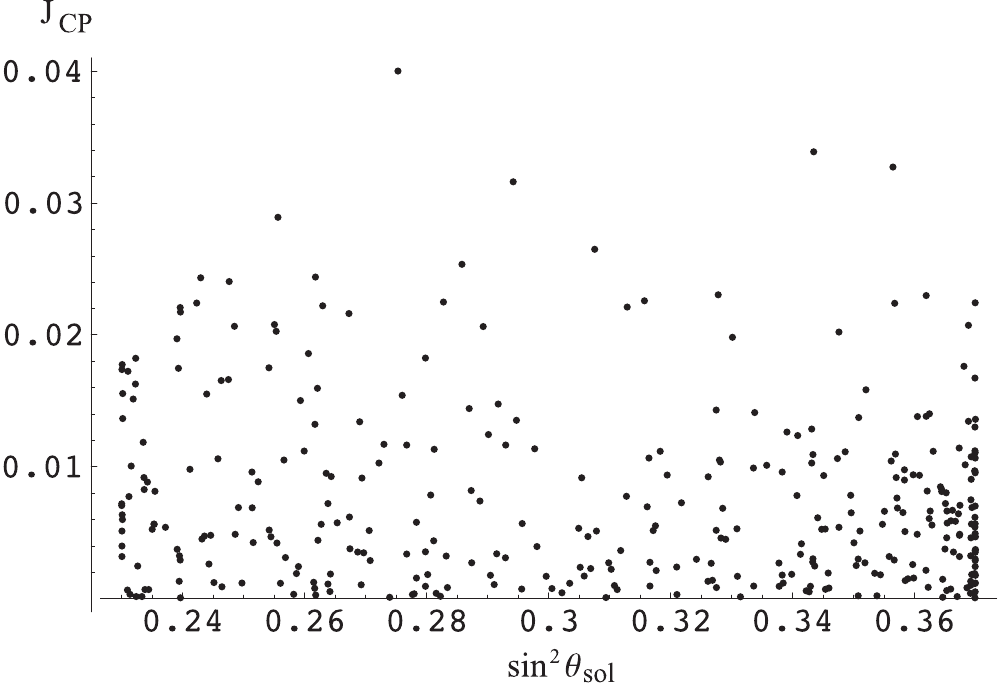}
\end{center}
\caption{Jarlskog parameter as a function of the solar angle.
}
\label{neutrinos}
\end{figure}

We diagonalise $M_{\chi^0}$ numerically and use $M_1\sim M_2\sim 1$ TeV.
Great care must be taken, as the elements of
$M_{\chi^0}$ may vary over ten orders of magnitude, and the eigenvalues
themselves over as much as twenty orders of magnitude.
Our calculations are carried out using a forced
minimal precision of fifty decimals.
The errors due to the lack of precision in
this case appear farther than eight places behind the
decimal point for the neutrino masses.
The diagonalising matrix $\mathcal{N}$, with
$\mathcal{N}^\ast M_{\chi^0} \mathcal{N}^{-1} =
\mathrm{diag}(m_{\chi^0_i},m_{\nu_j})$,
has the following general form
\begin{equation}
\mathcal{N}=\left(\begin{array}{cc}
\zeta&N_\chi\\V_\nu^T&\bar\zeta^T\end{array}\right).
\end{equation}
Here $\zeta,\bar\zeta \ll 1$ denote $8\times3$ matrices that can be
determined perturbatively, see \textit{e.g.}
\cite{Hirsch:2000ef}.
Our interest lies in the matrix $V_\nu$, the neutrino mixing
matrix.
Using the canonical notation for the neutrino mixing matrix
\cite{Hirsch:2000ef},
$U=V_{\nu}\ \cdot\mathrm{diag} (1,e^{i\phi_1},e^{i\phi_2})$ and
$c/s_{\ ij} = \cos/\sin\ \theta_{ij}$,
\begin{equation}
V_\nu= \left(\begin{array}{ccc}
c_{12}c_{13} & s_{12}c_{13} & s_{13}e^{-i\delta}\\
-s_{12}c_{23}-c_{12}s_{23}s_{13}e^{i\delta} &
c_{12}c_{23}-s_{12}s_{23}s_{13}e^{i\delta}  & s_{23}c_{13}\\
s_{12}s_{23}-c_{12}s_{23}s_{13}e^{i\delta}  &
-c_{12}s_{23}-s_{12}c_{23}s_{13}e^{i\delta} & c_{23}c_{13}
\end{array}\right),\quad
\end{equation}
we can extract the mixing angles as follows:
\begin{equation}
\nonumber
\sin\theta_{13}=\left|V_\nu^{13}\right|,
\ \tan\theta_{12}=\left|{V_\nu^{12}\over V_\nu^{11}}\right|,
\ \tan\theta_{23}=\left|{V_\nu^{23}\over V_\nu^{33}}\right|.
\end{equation}
One can also extract the CP violating Dirac phase $\delta$
\cite{Hirsch:2002tq}:
\begin{equation}
|\delta|=\sin^{-1}\left(\left|{8\,\mathrm{Im}(V_\nu^{21}V_\nu^{\ast
22}V_\nu^{12}V_\nu^{\ast 11}) \over
                                \cos\theta_{13}\sin 2\theta_{13}\sin
2\theta_{12}\sin 2\theta_{23}}\right|\right),
\end{equation}
where the Jarlskog invariant $J_{CP}$ \cite{Jarlskog:1985ht} of the 
neutrino sector is
given by:
\bea
J_{CP}=|\mathrm{Im}(V_\nu^{21}V_\nu^{\ast 22}V_\nu^{12}V_\nu^{\ast 11})|
=|\mathrm{Im}(V_\nu^{31}V_\nu^{\ast 33}V_\nu^{13}V_\nu^{\ast 11})|
=|\mathrm{Im}(V_\nu^{23}V_\nu^{\ast 22}V_\nu^{32}V_\nu^{\ast 33})|.
\label{eq:J_CP}
\eea
In the quark sector, the $J$ values in the SM are known to be
$J\sim 10^{-5}$.
For points satisfying all the constraints from the scalar sector,
we then apply the following experimental constraints concerning
the neutrino sector \cite{Mohapatra:2004ht}:
\bea
&&\sin^22\theta_{23}\ge 0.89,\;\;
\sin^2\theta_{13}\le 0.047,\;\;
\sin^2\theta_{12}\simeq 0.23 - 0.37,\nonumber\\
&&\Delta m_{atm}^2 \simeq 1.4\times 10^{-3} \mathrm{eV}^2 - 3.3\times
10^{-3} \mathrm{eV}^2,\nonumber\\
&&\Delta m_{sol}^2 \simeq 7.3\times 10^{-5} \mathrm{eV}^2 -
9.1\times 10^{-5} \mathrm{eV}^2.
\eea
In Fig. \ref{neutrinos} we show Jarlskog invariant
in Eq. (\ref{eq:J_CP})
for a sample of 350 points in the parameter space satisfying all
the scalar and neutrino sector constraints.
It is seen that $J$ is less than around 0.04.
These values open the possibility
of detecting the CP violation in the leptonic sector through
neutrino oscillations, see
\cite{CPobservability,Gonzalez-Garcia:2001mp} and references therein.

\section{Fermion Electric Dipole Moments and CP violation in the kaon
system: $\epsilon_K$}
\label{edms}

Electric dipole moments represent  a challenge for supersymmetric theories.
It is known that the MSSM predicts too large EDMs by about three orders of
magnitude for scalar fermion masses close to the current experimental
bounds ($\cal O$(100 GeV)) and CP violating phases of ${\cal O}(1)$
\cite{Ellis:1982tk}.
There are at present three solutions to this problem.
One is to assume that supersymmetric phases are not of order unity, but
rather of ${\cal O}(10^{-2} -10^{-3})$ \cite{Ellis:1982tk}.
The second possibility is that the spectrum of the supersymmetric partners
of quarks and leptons  is heavy, \textit{i.e.} of ${\cal O}(3~\rm TeV)$ 
or more
\cite{Kizukuri:1991mb}, and out of reach of the LHC.
The third possibility is that there are internal
cancellations among the different components of the neutron EDM (the
chargino and gluino contributions in particular) which can reduce the
magnitude of the neutron EDM \cite{Ibrahim:1997gj}. Analyses 
have demonstrated that these cancellations are very difficult to 
achieve \cite{Falk:1999tm}.
Finally attempts to set the flavor diagonal CP violation parameters 
to zero, but to allow CP violation through off diagonal elements in 
the scalar fermion mass matrices lead to too large EDMs, and further 
constraints must be imposed  \cite{Barbieri:1995tw}.
All these solutions are in effect fine tuning, either for the scalar fermion
masses, or for the phases, or for part of the parameter space.

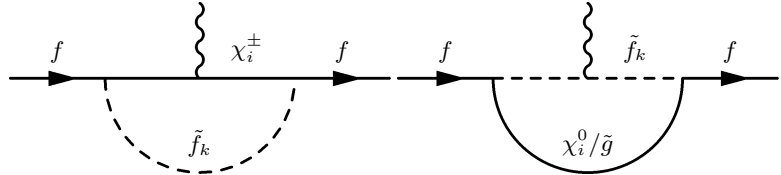
\begin{figure}[!htb]
\begin{center}
\begin{fmfgraph*}(50,20)
\fmfpen{thin}
\fmftop{photon}
\fmfleft{quarkIn}
\fmfright{quarkOut}
\fmf{quark,label=$f$,label.side=left}{quarkIn,v1}
\fmf{plain}{v1,v2}
\fmf{plain,label=$\chi_i^\pm$,label.side=left}{v2,v3}
\fmf{quark,label=$f$,label.side=left}{v3,quarkOut}
\fmffreeze
\fmf{dashes,right,label=$\tilde f_k$,label.side=left}{v1,v3}
\fmf{photon}{v2,photon}
\end{fmfgraph*}
\begin{fmfgraph*}(50,20)
\fmftop{photon}
\fmfleft{quarkIn}
\fmfright{quarkOut}
\fmf{quark,label=$f$,label.side=left}{quarkIn,v1}
\fmf{dashes}{v1,v2}
\fmf{dashes,label=$\tilde f_k$,label.side=left}{v2,v3}
\fmf{quark,label=$f$,label.side=left}{v3,quarkOut}
\fmffreeze
\fmf{plain,right,label=$\chi^0_i/\tilde{g}$,label.side=left}{v1,v3}
\fmf{photon}{v2,photon}
\end{fmfgraph*}
\caption{$\strut^{\strut}$
The loop contributions to fermion EDMs.
}
\label{feyEDM}
\end{center}
\end{figure}

The EDMs and $\epsilon_K$  in our model arise from loop contributions
and are straightforward to calculate.
The definition of the  EDM $d_f$ for a spin-${1\over 2}$ particle is
\begin{equation}
\mathcal{L}_I=-{i\over 2} d_f \bar\Psi \sigma_{\mu\nu}\gamma_5\Psi F^{\mu\nu},
\end{equation}
and the general interaction Lagrangian between two
fermions ($\bar\Psi$, $\Psi$) and a scalar ($\chi$) containing CP violation is
\begin{equation}
-\mathcal{L}_{int} = \sum_{ik} \bar\Psi_f\left(K_{ik}{1-\gamma_5\over 2}
   +L_{ik}{1+\gamma_5\over 2}\right)\Psi_i \chi_k +\mathrm{h.c.}.
\end{equation}
This gives us the one loop EDM as
\bea
\label{edm:Df}
d_f &=& \sum_{ik} {m_i\over(4\pi)^2 m_k^2}\mathrm{Im}(K_{ik}L_{ik}^\ast)\left[
Q_i A\left({m_i^2\over m_k^2}\right) +Q_k B\left({m_i^2\over 
m_k^2}\right)\right] ,\\
&&A(r)={1\over 2(1-r)^2}\left(3-r+{2\log r\over 1-r}\right) ,\\
&&B(r)={1\over 2(1-r)^2}\left(1+r+{2r\log r\over 1-r}\right).
\eea
Where $Q_i$, $Q_k$ are the charges of the fermion and scalar respectively.
Clearly we need $\mathrm{Im}(K_{ik}L^\ast_{ik})\neq 0$ for there to be
a nonzero contribution.
There are three different contributions depending on which particles
are running in the loop: chargino, neutralino or gluino.
As mentioned above, small EDMs are achieved if these contributions
cancel out.
{}From (\ref{edm:Df}) the other two common solutions are also easily
understood, since increasing the squark mass $m_k$ or suppressing
$\mathrm{Im}(K_{ik}L^\ast_{ik})\neq 0$ both yield a small $d_f$.

The CP-violating parameter
$\epsilon_K=\mathrm{Im}(\mathcal{M}_{K\bar K}/\Delta m_K)$ receives
no contribution from standard model processes in our model and the only
contribution to $\mathcal{M}_{K\bar K}$ comes from a chargino loop.
Compared to the EDMs, the calculation of the kaon oscillation loop is
more involved because there are non-perturbative hadronic states
in the process.
We use the vacuum insertion approximation (VIA) \cite{Branco:1999fs}
whereby the matrix element
$\langle K | (\bar s \Gamma d)(\bar s \Gamma d)|K\rangle$ appearing in
the loop calculation is reduced to two nonzero contributions
($\mathbf{V}_1$ and $\mathbf{V}_2$) which can be measured from kaon decays.

\begin{figure}[!htb]
\begin{center}
\begin{fmfgraph*}(60,25)
\fmfpen{thin}
\fmfleft{dIn,sIn}
\fmfright{sOut,dOut}
\fmf{quark,label=$s$,label.side=left}{v1,sIn}
\fmf{quark,label=$d$,label.side=left}{dIn,v2}
\fmf{quark,label=$s$,label.side=left}{v3,sOut}
\fmf{quark,label=$d$,label.side=left}{dOut,v4}
\fmf{quark,tension=.5,label=$\chi^-_l$,label.side=left}{v4,v1}
\fmf{quark,tension=.5,label=$\chi^-_k$,label.side=left}{v2,v3}
\fmffreeze
\fmf{dashes,tension=.5,label=$\tilde q_i$,label.side=left}{v1,v2}
\fmf{dashes,tension=.5,label=$\tilde q_j$}{v4,v3}
\end{fmfgraph*}
\begin{fmfgraph*}(60,25)
\fmfpen{thin}
\fmfleft{dIn,sIn}
\fmfright{sOut,dOut}
\fmf{quark,label=$s$,label.side=left}{v1,sIn}
\fmf{quark,label=$d$,label.side=left}{dIn,v2}
\fmf{quark,label=$s$,label.side=left}{v3,sOut}
\fmf{quark,label=$d$,label.side=left}{dOut,v4}
\fmf{dashes,tension=.5,label=$\tilde q_i$}{v1,v4}
\fmf{dashes,tension=.5,label=$\tilde q_j$,label.side=left}{v2,v3}
\fmffreeze
\fmf{quark,tension=.5,label=$\chi^-_l$}{v2,v1}
\fmf{quark,tension=.5,label=$\chi^-_k$,label.side=left}{v3,v4}
\end{fmfgraph*}
\caption{
The loop contributions to $\epsilon_K$.
}
\label{feyKaon}
\end{center}
\end{figure}
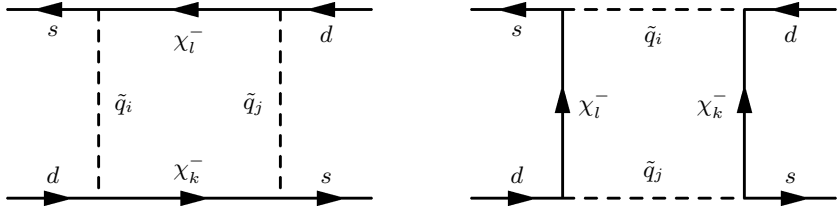

Defining the quark-squark-chargino interaction as
\begin{equation}
-\mathcal{L}_{q\tilde q\chi} =\bar q_i \left(V_{ijk}^L\gamma_L + 
V_{ijk}^R\gamma_R \right)\tilde q_j\chi_k + \mathrm{h.c.},
\end{equation}
and introducing the following notation:
\bea
W_{ijkl}^1& = &V_{sil}^{L\ast}V_{djl}^{L}V_{sjk}^{L\ast}V_{dik}^{L} + 
(L \leftrightarrow R) \nonumber ,\\
W_{ijkl}^2& = &V_{sil}^{L\ast}V_{djl}^{L}V_{sjk}^{R\ast}V_{dik}^{R} + 
(L \leftrightarrow R) \nonumber ,\\
W_{ijkl}^3& = &V_{sil}^{L\ast}V_{djl}^{R}V_{sjk}^{L\ast}V_{dik}^{R} + 
(L \leftrightarrow R) \nonumber ,\\
W_{ijkl}^4& = &V_{sil}^{L\ast}V_{djl}^{R}V_{sjk}^{R\ast}V_{dik}^{L} + 
(L \leftrightarrow R) \nonumber ,
\eea
we can write the matrix element as
\bea
\mathcal{M}_{K\bar K}&=& {i\over (2\pi)^4} {1\over 16\pi^2} 
\displaystyle\sum_{ijkl}
\left\{ {2\over m_i^2} \mathbf{I}_{ijkl}^1
\left[W_{ijkl}^1 {2\over 3} \mathbf{V}_2+W_{ijkl}^2
\left({1\over 3}\mathbf{V}_1-{1\over 
2}\mathbf{V}_2\right)\right]\right.\nonumber\\
&+& \left.{M_l M_k \over m_i}^4 \mathbf{I}_{ijkl}^2
\left[W_{ijkl}^3{5\over 12}\mathbf{V}_1+W_{ijkl}^4
\left(-{1\over 2}\mathbf{V}_1+{1\over
12}\mathbf{V}_2\right)\right]\right\}.
\label{element}
\eea
Here $\mathbf{I}^i$ are lengthy expressions arising from the loop 
integrals depending on the masses of the particles in the loop. 
Denoting $R_{ab}\equiv (m_a^2-m_b^2)/m_i^2$, they are
\bea
\mathbf{I}_{ijkl}^1&=&R_{lj}^{-1}\left\{R_{jk}^{-1}\left[
R_{ik}^{-1}\left({(1-R_{ik})^2\over 2}\left(\log(1-R_{ik})-{1\over 
2}\right)+{1\over 4}\right)\right.\right.
\nonumber\\
&&-\left.\left.R_{ij}^{-1}\left({(1-R_{ij})^2\over 
2}\left(\log(1-R_{ij})-{1\over 2}\right)+{1\over 4}\right)
\right] - R_{lk}^{-1}\bigg[j\to l\bigg] \right\},
\\
\mathbf{I}_{ijkl}^2&=&R_{lj}^{-1}\Big\{R_{jk}^{-1}\Big[
R_{ik}^{-1}(1-R_{ik})\log(1-R_{ik})-R_{ij}^{-1}(1-R_{ij})\log(1-R_{ij})\Big]
\nonumber\\
&&-R_{lk}^{-1}\Big[
R_{ik}^{-1}(1-R_{ik})\log(1-R_{ik})-R_{il}^{-1}(1-R_{il})\log(1-R_{il})
\Big]\Big\}.
\eea

It is easily seen how these expressions simplify when using the mass
insertion approximation (\textit{i.e.} $i=j$).
Since the model provides us with a full squark spectrum, we
prefer to use the full expressions in Eq. (\ref{element}).
The VIA coefficients are given as follows \cite{Branco:1999fs}:
\begin{equation}
\mathbf{V}_1= {f_K^2 m_K^4\over 2 m_K (m_s + m_d)^2},~~ 
\mathbf{V}_2={f_K^2 m_K^2 \over 2 m_K},~~ f_K \simeq 160~\mathrm{MeV}.
\end{equation}

\begin{figure}[tpb]
\begin{center}
\includegraphics[width=8truecm]{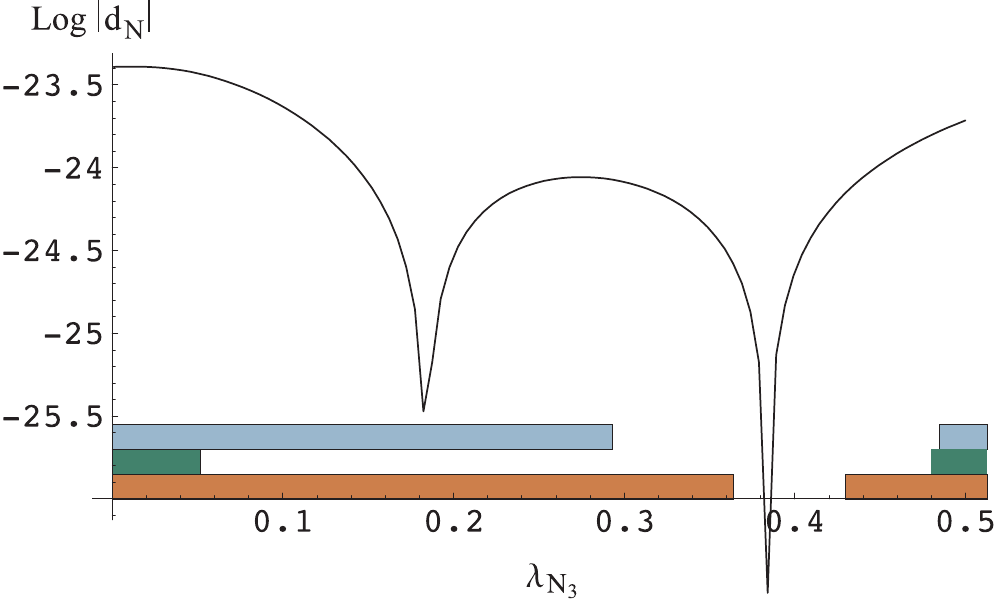}
\end{center}
\caption{
$\lambda_{N_3}$ effect on neutron EDM. The bars show 
excluded ranges due to problems with vacuum stability (top), charged 
Higgs mass limits (middle), and neutrino sector constraints 
(bottom).
}
\label{EDM}
\end{figure}

\begin{figure}[tpb]
\begin{center}
\includegraphics[width=8truecm]{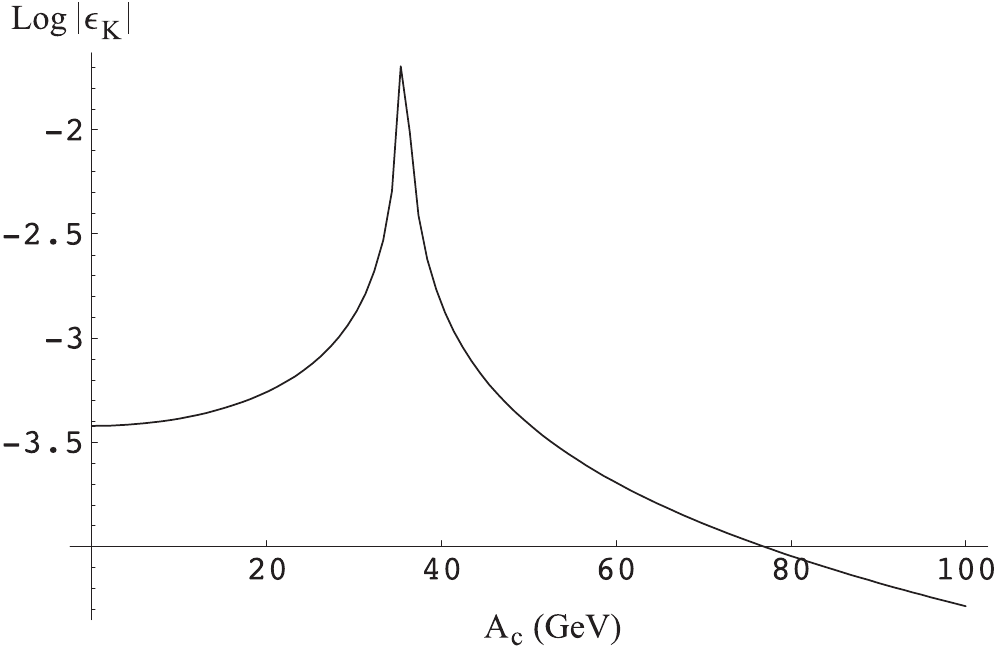}
\end{center}
\caption{
$\epsilon_K$ as a function of 
the trilinear coupling 
$A_c$.
}
\label{eps}
\end{figure}

Using these results, our method to search for viable points in the 
parameter space
has been to find first points
which 
satisfy all the Higgs mass and neutrino sector constraints.
 From these points we begin a random walk in parameter space,
slightly varying the parameter values for each step, checking the 
EDMs and discarding
such steps as do not bring us closer to the experimentally acceptable values.
There is a set of parameters ($A_u,\ M_3,\ M_U^{11}$) which enters 
exclusively in the calculation of the EDM.
Suppression of the EDM can be achieved by increasing these mass parameters or,
alternatively, one can find values for which cancellation between the 
different loop contributions occur.
In general, varying only these parameters can lead to undesirably large 
mass parameter values ($>\mathcal{O}$(3 TeV). In such cases
we vary all the parameters relevant to the EDM. Since the same 
parameters affect other
sectors of our model as well, we have to check, at each random step,
the various experimental constraints.
In Fig. \ref{EDM} we show the effect of changing $\lambda_{N_3}$
on the neutron EDM. The  curve shows two values
for $\lambda_{N_3}$ where cancellations between the contributions to 
the EDM occur.
It is seen that only one of the two dips in the curve satisfies 
all the required constraints.
After this we try, again by randomly
varying parameters, to find experimentally acceptable  $\epsilon_K$.
This process is easier than for the EDMs, since the set of parameters
($A_c,\ M_Q^{22},\ M_U^{22}$), which only affects the value of $\epsilon_K$,
is sufficient for reaching acceptable $\epsilon_K$ values. In Fig. \ref{eps} the behaviour of
$\epsilon_K$, for the same parameter point as in Fig. \ref{EDM}, is shown
as a function of one of the trilinear couplings, $A_c$. A clear peak
where we find the experimentally allowed value can be seen. For the
range shown, all the scalar and neutrino sector constraints,
as well as EDMs, remain viable.

We find that satisfying EDMs and kaon sector CP violation is possible
without contstraining the phases,
though quite restrictive, in the model.
The reason why satisfactory values are found is due to the structure of the
parameter space. There are variables, which affect EDMs or $\epsilon_K$,
but which do not affect the other observables.

\section{Conclusions and 
discussion}
\label{conclusion}

Violation of CP symmetry is well 
established, while neutrino masses
make violation of R-parity 
attractive.
We have considered here a model in which both R-parity 
and
CP-symmetry break spontaneously.
Our model contains in addition 
to the MSSM fields
only the three right-handed neutrinos $N_i$ and 
a singlet field $S$,
which are both needed to spontaneously break CP 
and R-parity.
We have shown that
experimentally viable neutrino and 
Higgs boson masses can be obtained,
and that CP is violated in the 
neutrino sector.
We explored the parameter space requiring that EDM 
bounds and experimental 
results on the kaon system 
(especially
$\epsilon_K$) are satisfied. 
In \cite{Hugonie:2003yu} 
solutions for models satisfying both constraints in 
 an R-parity 
conserving model were discussed.
In our case, there are more 
parameters in relevant sectors, like
 the chargino mass matrix,
which make the bounds mentioned easier to
 fulfill. 
 
 While this 
paper is dedicated to the presentation of the model, and 
constraints 
on its parameter space, two immediate consequences arise for
low 
energy phenomenology. In the leptonic sector, we predict a 
measurable CP violation, with the Jarlskog invariant 
below
$0.04$.
In the 
scalar sector, we predict reduced $ZZh_i$ couplings (with 
$h_i$ any 
neutral Higgs boson) compared
to the SM, and a Higgs strahlung 
production less frequent than in 
the SM. As well, the model favors 
a
lightest Higgs mass less than its  SM value, $m_H < 114$ GeV.
With our 
particle content, it is kinematically possible to produce
several 
neutral Higgses within the LHC mass reach.

In the present paper we 
have not studied the B-meson sector in detail, and we
 expect that 
modifications for the quark sector are needed before we can
agree with the experimental results.
Such modifications are achieved {\it e.g.}
by 
using a model where the Higgs sector leads to a complex CKM
matrix \cite{Doff:2006rt} or by
adding vector quarks 
\cite{Branco:2006pj}.
However, the parameters involved in the 
calculations of the neutrino
sector and calculations involving quarks 
are largely disconnected,
as we have seen from our kaon sector 
results.
Thus the results of the present model concerning the Higgs 
and neutrino 
sectors are not expected
to change qualitatively.
The 
work along these directions is in progress 
\cite{FHR}.

\begin{acknowledgement}

This work is supported by the 
Academy of Finland
(Project number 104368 and 115032) and by NSERC of 
Canada (0105354).

\end{acknowledgement}

\end{fmffile}
\end{document}